\ifx\pdftexversion\undefined
    \documentclass[10pt]{article}
\else
    \documentclass[10pt]{article}
\fi
\newcommand{\degr}{\mbox{$^\circ$}}%

\usepackage{times}

\ifx\pdftexversion\undefined
  \usepackage{graphicx}
\else
  \usepackage[pdftex]{graphicx}
\fi

\newcommand{\LkHa}{LkH$\alpha$ }

\newenvironment{sciabstract}{%
\begin{quote} \bf}
{\end{quote}}

\newcounter{lastnote}
\newenvironment{scilastnote}{%
\setcounter{lastnote}{\value{enumiv}}%
\addtocounter{lastnote}{+1}%
\begin{list}%
{\arabic{lastnote}.}
{\setlength{\leftmargin}{.22in}}
{\setlength{\labelsep}{.5em}}}
{\end{list}}

\title{Laser Guide Star Adaptive Optics Imaging Polarimetry of Herbig Ae/Be Stars}

\author{
Marshall D. Perrin$^{1,6*}$, James R. Graham$^{1,6}$, Paul Kalas$^{1,6}$, James P. Lloyd$^{2,6}$,\\
Claire E. Max$^{3,6}$, Donald T. Gavel$^{5,6}$, Deanna M. Pennington$^{3,6}$, Elinor L. Gates$^{4,6}$\\
\\
\normalsize{$^{1}$Astronomy Department, University of California Berkeley, Berkeley CA 94720}\\
\normalsize{$^{2}$Astronomy Department, California Institute of Technology, 1201 East California Blvd, Pasadena CA 91125}\\
\normalsize{$^{3}$Lawrence Livermore National Laboratory, 7000 East Avenue, Livermore CA 94550}\\
\normalsize{$^{4}$UCO/Lick Observatories. P.O. Box 85, Mount Hamilton CA 95140}\\
\normalsize{$^{5}$Laboratory for Adaptive Optics, University of California Santa Cruz, }\\
\normalsize{1156 High Street, Santa Cruz CA 95064}\\
\normalsize{$^{6}$NSF Center for Adaptive Optics, University of California Santa Cruz, }\\
\normalsize{1156 High Street, Santa Cruz CA 95064}\\
\\
\normalsize{$^\ast$To whom correspondence should be addressed; E-mail:  mperrin@astro.berkeley.edu}
}

\date{}

\begin{document} 




\maketitle


\begin{sciabstract}

We have used laser guide star adaptive optics and a near-infrared
dual-channel imaging polarimeter to observe light scattered 
in the circumstellar environment of
Herbig Ae/Be stars on scales of 100-300 AU.  We discover a strongly
polarized, biconical nebula 10 arcseconds in diameter (6000 AU) around
the star \LkHa 198, and also observe a polarized jet-like feature
associated with the deeply embedded source \LkHa 198-IR.  The star
\LkHa 233 presents a narrow, unpolarized dark lane consistent with an
optically thick circumstellar disk blocking our direct view of the
star.  These data show that the lower-mass T Tauri and 
intermediate mass Herbig Ae/Be stars share a common evolutionary sequence.

\end{sciabstract}

Diffraction-limited optical and infrared astronomy from the ground
requires adaptive optics (AO) compensation to eliminate atmospheric
wavefront disturbances. Bright stars may be used as wavefront
references for this correction, but most astronomical targets lack
nearby guide stars. AO observations of these targets from the ground
can only be accomplished using artificial laser guide stars (LGS)
\cite{1994OSAJ...11..263H}.

Herbig Ae/Be stars are young stars with masses between 1.5 and 10
times that of the sun; they are the intermediate-mass counterparts of
the more common T Tauri stars.  Excess infrared and millimeter
emission shows that Herbig Ae/Be stars are associated with abundant
circumstellar dust \cite{hil92}.  Visible and near-infrared (NIR)
light scattered from dust is typically polarized perpendicular to the
scattering plane \cite{1987ApJ...317..231B}, making polarimetry a
useful tool for probing the distribution of this material
\cite{2000prpl.conf..247W}.  While Herbig Ae/Be stars are
intrinsically very luminous, many are so distant or extincted that
they are too faint to act as their own wavefront references and thus
require LGS AO.  

The Herbig Ae/Be stars \LkHa 198 and \LkHa 233 (Table S1) were observed on 2003
July 22 at the 3-m Shane telescope at Lick Observatory (Fig. S1) using the
Lawrence Livermore National Laboratory LGS AO system\cite{max98} and
the Berkeley NIR camera IRCAL\cite{Lloyd00}.  The atmospheric seeing
was 0.8 arcseconds at 550 nm, and the AO-corrected wavefront produced
images with Strehl ratios $\approx 0.05-0.1$ at 2.1 microns and
full-width at half maximum resolution of 0.27 arcseconds \cite{SOM}.

IRCAL's imaging polarimetry mode utilizes a cryogenic LiYF$_4$
Wollaston prism to produce two simultaneous images of orthogonal
polarizations.  The sum of the two channels gives total intensity, and
the difference gives a Stokes polarization.  The dominant noise source
near bright stars in AO images is an uncorrected seeing halo. Because
this halo is unpolarized and thus vanishes in the difference image,
dual-channel polarimetry enhances the dynamic range in circumstellar
environments\cite{2000ApJ...540..422P}.  The observing techniques and
data reduction methods are based on \cite{kuhn01}.

\LkHa 198 is located at the head of an elliptical loop of optical
nebulosity extending 40 arcseconds to the star's southeast
\cite{li94}.  This complex region 600 parsecs distant includes a
molecular CO outflow \cite{canto84} and two Herbig-Haro jets
\cite{1995A&A...293..550C}, as well as two additional Herbig Ae/Be
stars in the immediate vicinity (V376 Cas and \LkHa 198-IR)
\cite{2000ApJ...531..494H} and a millimeter source (\LkHa 198-MM)
believed to be a deeply embedded protostar \cite{1994A&A...292L...1S}.
This proximity of sources requires high resolution observations to
disentangle the relationships between the various
components\cite{1997ApJ...485..213K}.  We discover a biconical nebula
$\sim$10 arcseconds in diameter (6000 AU), oriented north-south, with
polarization vectors concentric with respect to \LkHa 198 (Fig. 1).  The lobes of the reflection nebula are divided by a
dark, unpolarized lane that we interpret as a density gradient towards
the equatorial plane of a circumstellar disk and/or a flattened
envelope.  The north-south orientation indicates that \LkHa 198 is
unlikely to have created the giant elliptical nebula or the molecular
outflow to the southeast. However, it is consistent with \LkHa 198
being the source for the Herbig-Haro flow at a position angle (PA,
measured east from north) of 160\degr, a conclusion supported by the
extension of the polarized reflection nebula along this PA. However,
this is 20\degr\ away from the observed symmetry axis of the nebula,
requiring an inner disk axis tilted or precessing with respect to the
outer envelope.

The embedded source \LkHa 198-IR \cite{1993ApJ...417L..79L} is
detected in our $H$ and $K_s$ band data $5.5$ arcseconds from \LkHa
198 at PA=$5^\circ$.  A polarized, extremely blue, jet-like feature
extends $>$2 arcseconds (1200 AU) from \LkHa 198-IR at PA=105$\degr$
\cite{2002PASJ...54..969F}.  The polarization vectors of this apparent
jet are perpendicular to its long axis, indicating that \LkHa 198-IR
is the illuminating source, not \LkHa 198.  The jet appears to be half
of a parabolic feature opening toward the southeast, with its apex at
\LkHa 198-IR and its southern side partially obscured by the envelope
around \LkHa 198.  We suggest that the northwest side of a bipolar
structure around \LkHa 198-IR may be hidden at NIR wavelengths by the
dust indicated by millimeter observations \cite{1998A&A...336..565H}.
The orientations of circumstellar structures revealed by our images
confirm that \LkHa 198-IR is the best candidate for the origin of the
Herbig-Haro outflow to PA 135\degr, though based on geometrical
considerations we cannot entirely exclude the protostar \LkHa 198-MM.
By extension, the large elliptical nebula was most likely created by
outflow from \LkHa 198-IR, although we see it primarily in scattered
light from the optically much brighter \LkHa 198.

\LkHa 233 is an embedded A5e-A7e Herbig Ae/Be star which is
associated with a blue, rectangular reflection nebula 50 arcseconds in
extent, located in the Lac OB1 molecular cloud at 880 parsecs.  Our
imaging polarimetry reveals four distinct lobes bisected by a narrow,
unpolarized lane with PA$\approx$150$^\circ$ (Fig.  1).
The nebulosity around \LkHa 233 is extremely blue in the NIR, with its
east-west extent decreasing from 6 arcseconds (5300 AU) at $J$ and $H$
to 2 arcseconds (1800 AU) at $K_s$.  The orientation of the lobes
relative to the dark lane suggests that they are the limb-brightened
edges of a conical cavity in a dusty envelope illuminated by a highly
extincted star.  The radial extent of the dark lane (1000 AU) suggests
that it is associated with an equatorial torus characteristic of a
flattened infalling protostellar cloud \cite{1984ApJ...286..529T}, and
not a rotationally-supported disk.  

We find that the intensity peak of the star is shifted southwest
relative to the polarization centroid, with the displacement
increasing from 0.15 arcseconds at $K_s$ to 0.35 arcseconds at $J$.
This too indicates that the lane consists of optically thick
foreground material which has a flattened spatial distribution
consistent with a circumstellar disk or infalling protostellar cloud.
We do not see the star directly, but instead view a scattering surface
above the disk midplane. 

Our results are consistent with low-resolution, wide-field optical
imaging polarimetry\cite{aspin85}, which suggests a circumstellar
torus in the northwest-southeast direction perpendicular to a bipolar
reflection nebula.  Our interpretation is also consistent with the
existence of an optical [S~II] emission line jet \cite{corc98}
blue-shifted to the southwest.  The jet both bisects the nebulosity
and lies perpendicular to the proposed disk.  The geometry of the
reflection nebula indicates that the outflow is only poorly collimated
($\Delta \theta \approx 70\degr$) despite the apparently narrow jet
traced by optical forbidden line emission.  Because [S~II] line
emission arises preferentially in regions denser than the critical
density for this transition, the surface brightness distributions can
take on the appearance of a highly collimated jet, despite the fact
that the streamlines collimate logarithmically slowly
\cite{1994ApJ...429..781S}.

Observations of T Tauri stars have led to a general understanding of
the origin of solar-type stars \cite{1987ARA&A..25...23S}: The
fragmentation and collapse of an interstellar cloud creates a
self-gravitating protostar surrounded by a Keplerian accretion disk
fed by an infalling, rotationally-flattened envelope.  The disk
mediates the outflows common to low-mass young stellar objects, which
play a key role in the dispersal of the natal gas and dust.  

It has been hypothesized that the more massive Herbig Ae/Be stars and
the T Tauri stars form and evolve in similar manners
\cite{1993ApJ...418..414P}, but this remains controversial.  On large
spatial scales, the disk-like nature of the circumstellar matter
around Herbig Ae/Be stars is well established.  Flattened structures
around several sources have been resolved on 100 AU scales
\cite{1997ApJ...490..792M}, or have Keplerian kinematics
\cite{1997Natur.388..555M}.  However, the evidence seems to be
ambiguous on scales of tens of AU and below, with some authors arguing
for a spherical geometry \cite{1994ApJ...432..710D} and others
favoring disks \cite{2001A&A...371..186N}.  

\LkHa 198 and \LkHa 233 are both classified as Hillenbrand Group II
Herbig Ae/Be stars: they have infrared spectra that are flat or rising
towards longer wavelengths. This means that they are young stars that
may or may not possess circumstellar disks but do possess
circumstellar envelopes which are not confined to a disk plane.  We
observe such envelopes around both of our sources in the form of
centrosymmetrically polarized biconical nebulosities viewed
approximately edge-on to the midplanes.

We compared our observations with radiative transfer models computed
for a sequence of circumstellar dust distributions around T Tauri
stars \cite{2003ApJ...591.1049W,2003ApJ...598.1079W}.  These models
provide both total and polarized intensity images, which we convolved
with a model instrumental point spread function to match our observed
resolution (Fig. S2).  Bipolar outflow cavities in these models produce a
limb-brightened appearance at near-infrared wavelengths, with the
brightening stronger in polarized light than in total intensity.

For both \LkHa 233 and \LkHa 198, the peak polarization differs
between the two lobes. These asymmetries may indicate the sign of the
inclination of each object. In Whitney's model envelopes, which use 
dust grain properties fit to an extinction curve for the Taurus molecular
cloud, the closer lobe was brighter overall but had lower fractional
polarization by several percent.  For \LkHa 198, the polarization in
the northern lobe of the bipolar nebulosity is generally 10-15\% lower
than the southern lobe at all wavelengths (20\% vs. 35\% at $H$, for
instance), suggesting that the northern lobe is oriented towards us.
For \LkHa 233, the southwest lobe is 14\% polarized on average versus
26\% for the northeast lobe. This indicates that the southern lobe is
facing us, consistent with the the blueshift of the CO jet in that
direction, and the southwest shift of the intensity peak relative to
the polarization centroid.

\LkHa 233's limb-brightened appearance provides compelling evidence
for the presence of cavities swept out by bipolar outflow from the
star.  Cavity models with an opening angle of 30-40\degr\ seen at an inclination
of 80\degr\ reproduce both the observed polarization fraction
of 25-40\% and the higher degree of limb brightening seen in the near lobe. 

The absence of limb brightening may be evidence that \LkHa 198 lacks
polar cavities, or at most possesses very narrow ones.  The
detectability of limb brightening for a given angular resolution
depends on the opening angle between the symmetry axis and edge of the
cavity.  At our resolution, envelope models with cavity opening angles
greater than $\sim 20\degr$ predict detectable limb brightening, while
we observe for \LkHa 198 an opening angle of $45\degr$ without limb
brightening.  Thus our observations do not support the presence of 
evacuated cavities in the envelope of \LkHa 198.  The observed
morphology can instead be explained by the illumination of a
cavity-free, rotationally-flattened envelope by the central star; the
bipolar appearance would then arise from light escaping along the path
of least optical depth.  However, these cavity-free infalling envelope
models have opening angles which increase with wavelength, while we
observe a constant opening angle, suggesting a geometric rather than
optical depth origin for the observed morphology.  This discrepancy
may be resolvable by varying the dust particle properties.

Based on these observations, \LkHa 233 is the more evolved of the two
systems, with well-defined cavities swept out by bipolar outflow and
bisected by a very dark lane. \LkHa 198 is a less evolved system,
which is only in the early stages of developing bipolar cavities and
possesses lower extinction in the apparent disk midplane.

The observed circumstellar environments are consistent with the
rotationally-flattened infall envelopes models developed for T Tauri
stars, indicating that the process of envelope collapse has similar
phases, despite the large disparities in mass and luminosity between
these two classes of young stars. This morphological similarity leads
us to infer that the conservation and transport of angular momentum is
the dominant physical process for both classes of stars.  Alternate
formation pathways have been suggested for OB stars that invoke new
physical mechanisms, such as magnetohydrodynamic turbulence
\cite{2003ApJ...585..850M} or stellar mergers
\cite{1998MNRAS.298...93B}.  The Herbig Ae stars studied here appear
to be below the mass threshhold at which such effects become
important. 


\clearpage

\begin{figure}
\vskip -1in
\ifx\pdftexversion\undefined
    \includegraphics[scale=0.9]{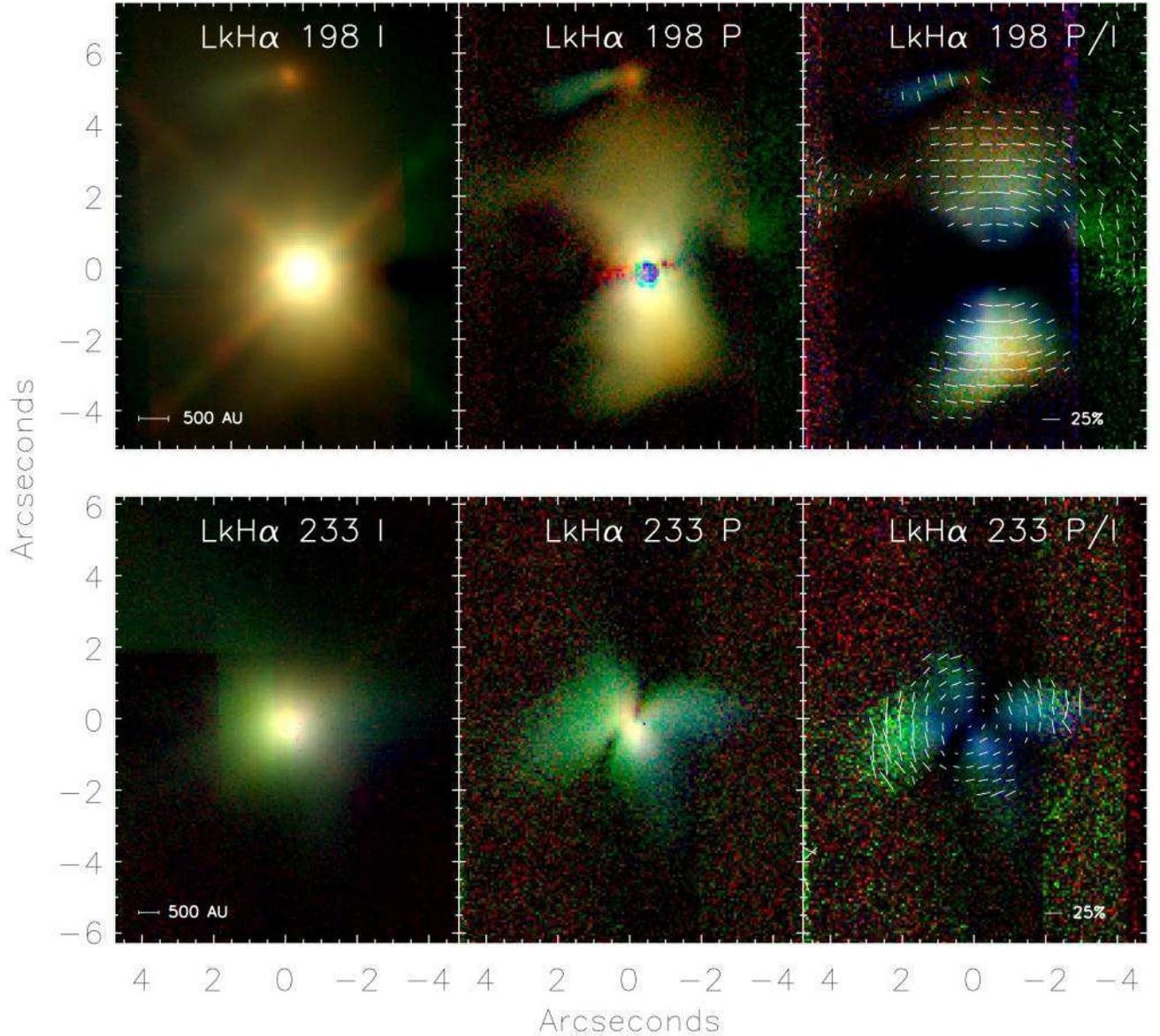}
    \else
     \includegraphics[scale=0.7]{haebes}
\fi
\caption{
 Three-color LGS AO mosaics of \LkHa 198 and
\LkHa 233. Plotted from left to right for each object are the total
intensity (Stokes $I$), the polarized intensity ($P = \sqrt{Q^2+U^2}$)
and the polarization fraction ($P/I$). $I$ and $P$ are displayed using
log stretches, while $P/I$ is shown on a linear stretch. Red is $K_s$
band (2.1 microns), green $H$ (1.6 microns), and blue $J$ (1.2
microns). Polarization vectors for $H$ band are overplotted on the
$P/I$ image; while the degree of polarization changes somewhat between
bands, the position angles do not vary much.  Integration times per
band were 960 s and 1440 s for \LkHa 198 and \LkHa 233, respectively.
The dimmest circumstellar features detected in our polarimetric
observations are approximately $1-2\times10^4$ fainter than the
stellar intensity peaks.  The IRCAL polarimeter is sensitive to
polarization fractions as low as a few percent, resulting in a
signal-to-noise ratio of 5-10 per pixel for the typical polarizations
of 15-40\% observed around our targets.  \label{polfig}}
\end{figure}

\clearpage

\textbf{LGS AO Imaging Polarimetry of Herbig Ae/Be Stars}\\
\textbf{Supporting Online Material}

\vskip0.25in
\textbf{Methods and Materials}

\vskip0.25in

The Lick Adaptive Optics system was developed at Lawrence Livermore
National Laboratory, and can operate in both natural and laser guide
star modes \cite{max98,2002SPIE.4494..336G}.  In the laser guide star
mode, the atmospheric wavefront reference is created by a laser tuned
to the sodium D2 line at 589 nm, which excites mesospheric sodium at
roughly 90 km altitude.  The 589 nm light is generated by a tunable
dye laser pumped by a set of frequency-doubled solid-state (Nd:YAG)
lasers. Typically, 11-14 W of average laser power is projected into
the sky with a pulse width of 150 ns and a pulse repetition rate of 13
kHz.  Laser guide star systems are insensitive to tip and tilt,
requiring a separate tip/tilt sensor using a natural guide star.  For
the observations presented here, the science targets served as their
own tip/tilt references.

The sodium guide star has an apparent size of ~ 2 arcseconds in 1 arcsecond
seeing and a magnitude which depends on the atmospheric sodium density, which
varies on all timescales from hourly to seasonally.  The sodium level was low
during July 2003, decreasing the magnitude of the guide star, and forcing the
adaptive optics system to operate at its lowest frame rate of 55 Hz.  As a
result, the Strehl ratios achieved were modest (S ~ 0.05-0.1) despite the good
atmospheric seeing.  Correspondingly the full-width at half-maximum (FWHM) of
the point spread function was larger than the FWHM of a diffraction limited
beam, 0.27 arcseconds versus 0.15 arcseconds respectively at 2.1 microns.

The science camera used with the Lick AO system is IRCAL \cite{Lloyd00}, which
has as its detector a $256^2$ pixel HgCdTe PICNIC array manufactured by Rockwell.
The observations presented in this paper used the standard astronomical $J$ (1.24
micron), $H$ (1.65 micron), and $K_s$ (2.15 micron) broad-band filters.  IRCAL's
plate scale, 0.0754 arcsec/pixel, was chosen to Nyquist sample the
diffraction-limited beam at $K_s$.  The imaging polarimetry mode of IRCAL
utilizes a cryogenic LiYF$_4$ (frequently called "YLF") Wollaston prism to
produce simultaneous images of orthogonal polarizations. YLF was chosen for its
excellent achromaticity throughout the near infrared. A rotating achromatic
half-wave plate mounted immediately before the camera entrance window modulates
the polarization, allowing measurement of both Stokes parameters $Q$ and $U$.

Each target was observed for the same amount of time
in $J$, $H$, and $K_s$, divided equally between Stokes $Q$ and $U$
observations. Typical exposures were 30-90 s in duration, with small
dithers performed every few exposures. Total integration time
was 1440 s per band for \LkHa 233, and 960 s per band for \LkHa 198.
The data were flat-fielded and bias-subtracted in the standard manner for
near infrared astronomical data. Sky background frames were obtained in 
polarimetric mode and subtracted from the data. However, the near infrared sky 
is nearly unpolarized so this step is not essential. The data from different dither 
positions were registered together via a Fourier transform cross-correlation code and 
stacked to produce mosaic Stokes $I$, $Q$, and $U$ images. 
These observing techniques and data reduction methods are based on \cite{kuhn01}.

The instrumental polarization bias was established through observations of
standard stars known to be unpolarized; From the derived bias (~ 2\% at a
position angle of -85\degr) we calculate the effective Mueller polarization
matrix for the instrument and apply the inverse of this matrix to the Stokes
mosaics to remove the bias.

\clearpage
\begin{table}[h!]
\caption[S1]{Target Summary}
\bigskip
\begin{tabular}{lrrrrrr}
\hline
{Object} & {$V$} & {Distance }  & {Spectral Type}& Luminosity & Mass  & {Time per band } \\
{} & {(magnitudes)} & {(parsec)}  & {}& ($L_\odot$) & ($M_\odot$) & {(s)} \\
\hline
\LkHa 198 & 	14.3 & 600 &   A5e-A7e   & 5.6 & 1.2 & 960 \\
\LkHa 233 & 	13.6 & 880 &   A5e    &  28.2 & 2.6 & 1440 \\
\hline
\end{tabular}
\bigskip
{\\
$V$ magnitude, distance, luminosity, and mass are from
\cite{hil92}. }

\end{table}


\clearpage

\begin{figure}
\ifx\pdftexversion\undefined
    \includegraphics[scale=0.5]{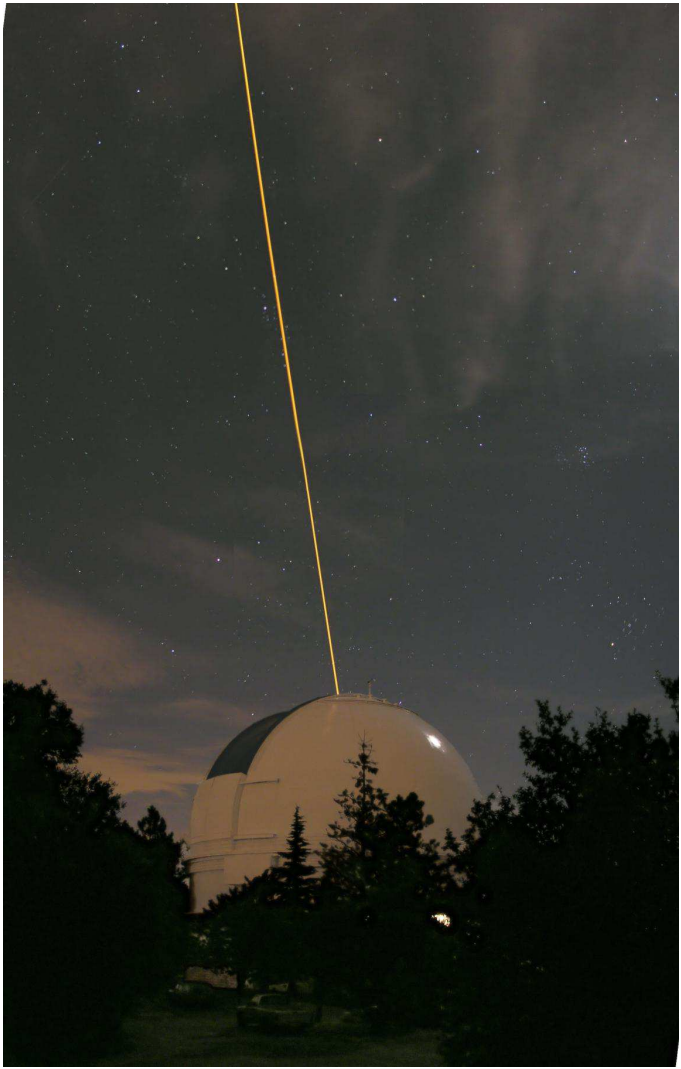}
\else
    \includegraphics[scale=1.2]{licklgs2b}
\fi
\caption{The Lick Observatory LGS AO system in operation on 2003 July
22. The laser beam is visible in Rayleigh scattered light for several
kilometers. The faint cirrus clouds illuminated by the Moon remained
outside our pointing direction and did not interfere with the observations. The
yellowish cast of the dome is due to the street lights of nearby San
Jose.  }

\end{figure}

\clearpage


\begin{thebibliography}{10}
\bibitem{SOM}
    Additional information on materials and methods is available as
    supporting material on Science Online.
\end{thebibliography}


\begin{scilastnote}
\item We are indebted to the Lick Observatory staff who assisted in
      these observations, including Tony Misch, Kostas Chloros, and
      John Morey, and also to the many individuals who have
      contributed to making the laser guide star system a reality. We
      thank Barbara Whitney for providing us with electronic versions
      of her models. Onyx Optics fabricated our YLF Wollaston prisms.
      This work has been supported in part by the National Science
      Foundation Science and Technology Center for Adaptive Optics,
      managed by the University of California at Santa Cruz under
      cooperative agreement No. AST-9876783; and also under the
      auspices of the U.S. Department of Energy, National
      Nuclear Security Administration by the University of California,
      Lawrence Livermore National Laboratory under contract No.
      W-7405-Eng-48. PK received additional support from the NASA
      Origins Program under grant NAG5-11769.  MDP is supported by a
      NASA Michelson Graduate Fellowship, under contract to the Jet
      Propulsion Laboratory (JPL). JPL is managed for NASA by the
      California Institute of Technology.  \end{scilastnote}

\noindent\textbf{Supporting Online Material}\\
www.sciencemag.org\\
Materials and Methods\\
Table S1\\
Figures S1, S2


\end{document}